\documentclass[1p,12]{elsarticle}
\usepackage{setspace}
\usepackage{lineno,hyperref}
\modulolinenumbers[5]

\usepackage{amsmath}
\usepackage{amssymb}
\usepackage{amsthm}
\usepackage{float}
\usepackage{xcolor}

\newcommand{\dt}{\frac{\textrm{d}}{¨\textrm{d}t}}

\begin{document}

\begin{frontmatter}

\title{Synchronization and attractors in a model simulating social jetlag}


\author{Flávia M. Ruziska \corref{mycorrespondingauthor}}
\cortext[mycorrespondingauthor]{Corresponding author}
\ead{flaviamayumi.rh@gmail.com}
\author{Iberê L. Caldas}
\address{Instituto de Física, Universidade de S\~{a}o Paulo, Rua do Mat\~{a}o, 1371, 05508-090 S\~{a}o Paulo, SP, Brazil}



%

\begin{abstract}
Much work has been done to investigate social jetlag, a misalignment between the biological clock and the social agenda caused by exposition to different light inputs, that causes several health issues. To investigate synchronization and attractors due to a sequence of light inputs, we introduce an extension of a model, previously used to describe jetlag caused by a single change in the light input. The synchronization to the light input is sensitive to the control parameters of the system and to the light input periods. Depending on the parameter set, the observed synchronization is with one or another successive light input. Most of the solutions have the period of a light input. However, for some parameters, we also observed higher period, chaotic solutions, and bistability.
\end{abstract}


\end{frontmatter}


\section{Introduction}

The circadian rhythms are 24-hours cycles produced by internal processes and entrained by external stimulus. The master clock of these cycles in mammals is the suprachiasmatic nucleus (SCN). A pathway from the retina to the suprachiasmatic nucleus (SCN) allows the synchronization of circadian rhythms to the solar days \cite{Reppert2002,10.3389/fphys.2019.00682}.  

Disruptions in the circadian rhythms can implicate several health issues on the short and on the long term. 
The disruption in the circadian rhythms which most people know is the jetlag. Jetlag occurs when someone travels through time zones, since different time zones have different light-dark cycles   
\cite{yamaguchi_mice_2013, lu_resynchronization_2016}). Jetlag causes effects like  impaired daytime function, general malaise and gastrointestinal disruption in the days immediately following travel \cite{lee2012jet}. Although these effects are unpleasant and harmful for some groups, there are also other circadian rhythms disruptions that affect many people generating health issues on the long term. 
For instance, shift-workers (e.g.  firemen, doctors on night shifts) usually suffer from disruptions  in the circadian rhythms. Empirical evidences indicate the association between these disruptions and negative
outcomes related to heart health, metabolic and gastrointestinal health, mental health, and cancer. \cite{James2017,om2003,oishi2005longitudinal,torquati2018shift}.  However, not only the shift-workers suffer from disruptions in the circadian rhythms. For example, as Smarr and Schirmer point out in \cite{smarr}, most students experience social jetlag correlated to decreased performance. Social jetlag is a misalignment between the social dynamic and the endogenous circadian rhythms of one individual. Several studies have been made in order to show the relation between social jetlag and other disorders. For instance, there are relations between social jetlag and depression \cite{sleep}; also, social jetlag is a risk factor for obesity and related diseases \cite{motaplos}. For these reasons it is important to study models that consider systematic disruptions in the circadian rhythms, and researchers are currently doing works in this area \cite{10.3389/fphys.2018.01529}. In this context, in this article we consider that the circadian rhythms are produced by endogenous oscillators that synchronize with the light input cycle.

From the dynamical system point of view, it is interesting to find a simple model that captures the main characteristics of a system. For that reason, here we adopted as base the model introduced in \cite{yamaguchi_mice_2013}, defined by four ODEs and yet able to reproduce the reported experimental data. In \cite{yamaguchi_mice_2013} it was experimentally shown that mice without arginine vasopressins (AVP) receptors, or mice in which the  arginine vasopressins (AVP) was pharmacology blocked, recovered faster from jetlag than regular mice. It is worth mentioning that the substance Arginine-vasopressin (AVP) is synthesized and released from nerve terminals in the central nervous system in the brain. This substance modulates several neuronal functions, for instance, arginine-vasopressin (AVP) is involved in the control of stress, cognitive behaviors and circadian rhythms \cite{Thomson2010}. 

In reference \cite{yamaguchi_mice_2013} the authors studied a scenario  similar to a jetlag, with only one shift in the light input. We, on the other hand, are interested in systematic shifts, that emulate the artificial light input of a shift-worker or of a person who suffers from social jetlag. Thus, we extend the model used in \cite{yamaguchi_mice_2013} for mice jetlag to investigate social jetlag. In fact, similarities between the SCN mamals have been discussed in the literature \cite{chang2001circadian}. We are assuming that changes on the artificial light input represent changes on the routine because nowadays people can control their artificial light input through electrical light. We are interested in understanding how the model solutions (attractors) change when this new type of light input is inserted. We also aim to understand the effect of the variation of the coupling parameter, which accordingly to \cite{yamaguchi_mice_2013} is related to arginine-vasopressin (AVP).     

Exploring the model in a large set of parameters and using techniques that allow us to measure numerically if a time-series is better synchronized with one stimulus or another, we find different scenarios of synchronization depending on the parameter. 
For most parameters  the solutions are period 1 (considering a Poincaré section in which we take one value per week). However, for some parameters, there are solutions of higher periods or chaotic, or even bistable solutions.

In section 2 we define the model; in section 3 we present the results first
treating one specific light input sequence; in section 4 we generalize our analyzes, we   present the results for a set of light inputs sequences and a range of the coupling parameter; finally in section 5 we discuss
the main results and highlight our conclusions.

\section{Model}

To describe the suprachiasmatic nucleus (SCN) behavior, that  allows the synchronization of circadian rhythms to the solar days \cite{Reppert2002,10.3389/fphys.2019.00682} , we use a model introduced in \cite{yamaguchi_mice_2013}.  In this model $\phi_0$ is the phase of the oscillator 0, which receives the light input $L(t)$, representing the cells in the SCN which receive light. $\phi_1$ and $\phi_2$ are the phases of oscillators 1 and 2, which are coupled with oscillator 0 and between themselves and represent two parts of the SCN. $\phi_3$ is the phase of oscillator 3, which represents a peripheral oscillator that receives the signal of SCN.  $K_{couple} = K_{AVP} + K_c$ is the strength of the coupling between the oscillators 1 and 2. For more details of the biological interpretation of the terms see supplementary material of  \cite{yamaguchi_mice_2013}. The model equations are:

\begin{align}
\dt \phi_0 &= \omega_0 + K_aZ(\phi_0,\varphi_a)L(t)\label{eq:1}\\
\dt \phi_1 &= \omega_1 + K_bZ(\phi_1,\varphi_b)h(\phi_0+0.05) + K_{couple}Z(\phi_1,0)h(\phi_2 +\alpha),\\
\dt \phi_2 &= \omega_2 + K_bZ(\phi_2,\varphi_b)h(\phi_0+0.05) + (K_{couple})Z(\phi_2,\alpha)h(\phi_1),\\
\dt \phi_3&=\omega_3 + K_dZ(\phi_3,\varphi_d)G(\phi_1,\phi_2),
\end{align}
where,
\begin{align}
&Z(\phi_l,\varphi_l) =-\sin[2\pi(\phi_l+\varphi_l)]\\  
&h  = \sin(\pi \phi /\theta), \quad 0 \le \mod(\phi,1)  \le \theta; \quad 0 \quad \textrm{ otherwise}\label{eq:5}. \\
&G(\phi_1,\phi_2)=\left(\frac{2+\sin(2\pi\phi_1)+\sin(2\pi\phi_2)}{4}\right)\\
&g_3 = \sin(2\pi \phi_3)\\
&\varphi_a = 0.1,\varphi_b=0.4,\varphi_d =-0.08, \alpha=0.35, \theta = 0.1\\
&k_{couple}  =K_{AVP} + K_c \\
&K_a = 8.0, K_b = 2.7, K_c = 0.6, K_d = 0.5.\\
&K_{AVP} = 0.7 \textrm{ represent regular mice } \\
&K_{AVP} = 0 \textrm{ represent modified mice without receptors of AVP} 
\end{align}
In \cite{yamaguchi_mice_2013} the authors showed that this model reproduces pretty well the experimental data they analyzed. Experimentally, the effect is when there are no receptors of arginine-vasopressin (AVP), or when this substance is blocked the mice recover faster from the jetlag. In the model, the parameter representing this substance is $K_{avp}$, and the authors verified that if the system is submitted to a light input shift (travel through timezone), it reaches the new equilibrium faster when $K_{AVP}$ (and consequently $K_{couple}$) is lower. It is worth remembering that in the model adopted $K_{couple} = K_{AVP} + K_c$ represents the coupling between the neurons, the boundaries of the interval in which we vary $K_{couple}$ are the extreme cases studied in \cite{yamaguchi_mice_2013}.  

In our work, instead of considering a light input with only one shift to emulate a jetlag scenario as in the original model, we consider periodic shifts in the light input, to emulate a scenario in which there is an abrupt change in the light input weekly. The idea is that this kind of light input is able to represent a sketch from the artificial light input which a shift-worker is used to, or, more generally, of a person who suffers from social jetlag is used to.

\begin{figure}[H]
	\centering
	\includegraphics[scale=0.7]{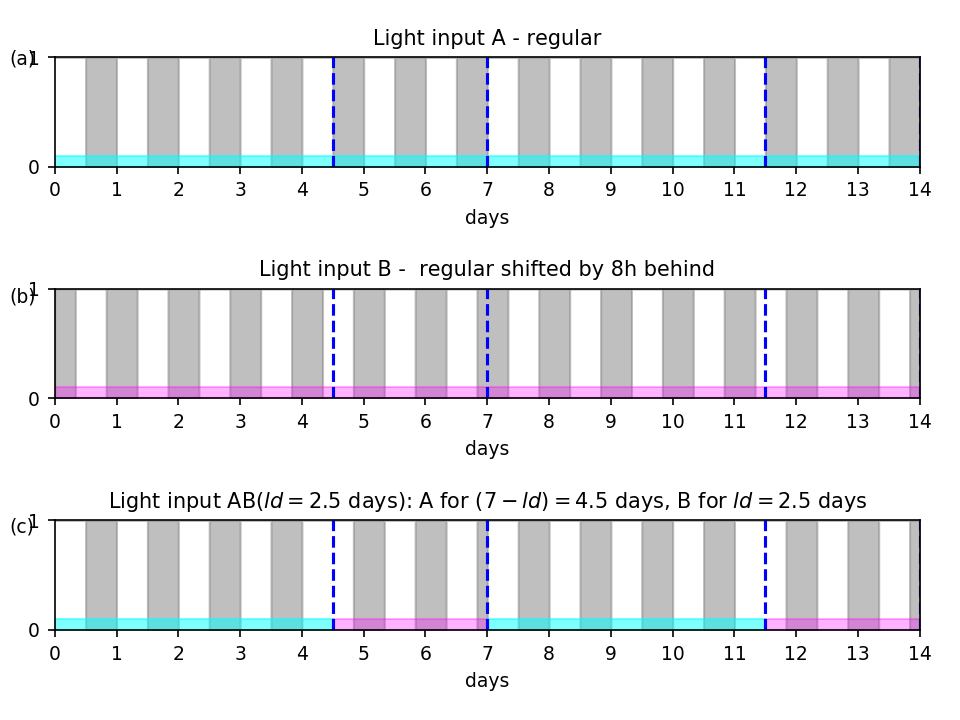}
	\caption{(a) Light input $A$ - regular light input, 12 hours of light (-12 to 0) followed by 12 hours of dark (0 to 12). (b) Light input $B$ 12 hours of light (-4 to 8) followed by 12 hours of dark (8 to -4), in other words, light input $A$ with a delay of 8h  (c) Light input $AB(l = 2.5)$, light input $A$ for 4.5 days followed by light input $B$ for 2.5 days.}
	\label{fig:jetlagsociallights}
\end{figure}

We divide the week in two parts: in the first part, the individuals receive the light input $A$; and in the second part they receive the light input $B$. Light inputs $A$ and $B$ are both regular, 12 hours of light followed by 12 hours of dark. However $B$ has a delay of 8h in relation to $A$, in other words, $B$ could be associated to a timezone 8h behind a timezone associated to light input $A$. The individual spends $l$ days in $B$; and (7 - $l$) days in  light input $A$. We name the light input through all the week light input $AB(l)$. In figure \ref{fig:jetlagsociallights} we show a scheme of the light input for the specific case $l=2.5$.

First we apply the model for a specific value of $l =2.5$.  Next we consider solutions for $l$ in the whole interval $[0,7]$ and parameter $k_{couple}$ in the interval $[0.6,1.3]$.  \\


\section{Results for light input with a shift in weekends}

In this subsection we analyze the specific case light input $AB(l = 2.5)$. We could also call this a week-weekend light input since in the week the light input is equal to light input $A$ and in the weekend is equal to light input $B$. 

\begin{figure}[H]
	\centering
	\includegraphics[scale = 0.7]{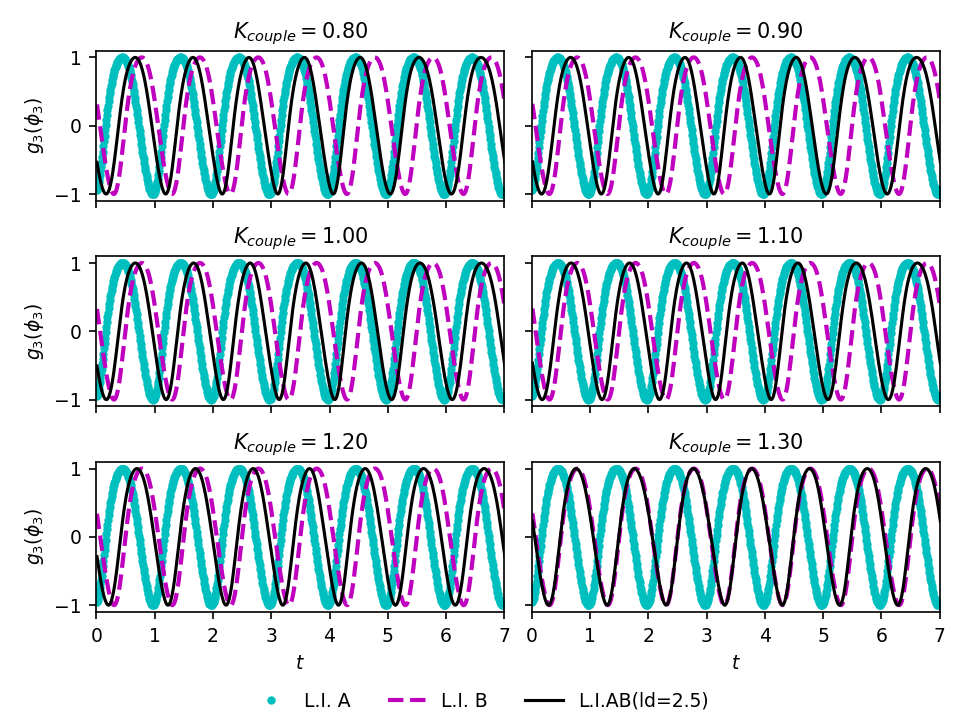}
	\caption{\label{fig:comparingfunctionsdirectly} The dotted cyan line corresponds to the output $g3(\phi_3)$ for the light input $A$, the dashed pink line to light for the light input $B$ and the black continuous line for the light input $AB(l=2.5)$. In each panel the value of the coupling parameter $K_{couple}$ is indicated.}
	
\end{figure}

In figure \ref{fig:comparingfunctionsdirectly} we show the  function $g_3(\phi_3)$ for  three different light inputs ($A$, $B$, $AB(l = 2.5)$). This function is defined as $g_3(\phi_3) = \sin(2 \pi \phi_3)$ where $\phi_3$ is the phase of the oscillator 3 of the model, which represents the peripheral oscillator that receives the signal of the SCN.  

Analyzing figure  \ref{fig:comparingfunctionsdirectly} we notice that as $K_{couple}$ increases the curve for $AB(l=2.5)$ gets similar to the curve for light input $B$; when $K_{couple} = 1.3$ they almost coincide.

The signal is better synchronized with the light input $B$ (weekend) than $A$ (week) for $K_{couple} = 1.3$. Since the model is highly nonlinear such result can be produced.

Aiming to evaluate numerically how the phase $\phi_3$ changes according different light inputs we introduce the distribution $z$ defined as 
\begin{equation}
z(\phi_3,\phi_3^\prime) =  e^{ i(\phi_3 \phi_3^\prime) } = e^{ i[2\pi(\phi_3 -\phi_3^\prime)] }  \label{eq:z}
\end{equation}
The first moment of $z(\phi_3,\phi_3^\prime)$ is given by
\begin{equation}
R(\phi_3,\phi_3^\prime)  = \frac{1}{N}\left|\sum_{i=1}^Nz_i\right|,\label{eq:R}
\end{equation}
which can be understood as a measurement of how much the two phases are synchronized; $N$ is the number of points used to evaluate te average. Finally, the mean angle $\theta$ of the first moment given by
\begin{equation}
\theta = Arg\left(\frac{1}{N}\sum_{i=1}^Nz_i\right),\label{eq:theta1}
\end{equation}
and it estimates the mean lag between $\phi_3$ and $\phi_3^\prime$.

Depending on our question we can choose different $\phi_3$ and $\phi_3^\prime$. To simplify the notation, we name the phase $\phi_3$ generated by the light input $A$ as $\phi_{3A}$; by the light input $B$  as $\phi_{3B}$; and by the light input $AB(l)$ as $\phi_{3AB}$. 

In the table \ref{tab:rtheta} we present the values for $R$ and $\theta$  fixing $\phi_3$ as $\phi_{3AB(l =2.5)}$, considering $\phi_3^{\prime}$ as $\phi_{3A}$ or $\phi_{3B}$ and the extreme values of $K_{couple}$ 0.6 and 1.3.

\small{
	\begin{table}[H]
		\caption{\label{tab:rtheta}Numerical comparison of $\phi_{3AB(l =2.5)}$ to $\phi_{3A}$ and $\phi_{3B}$ }
		\centering
		\begin{tabular}{c|c|c|c|c}
			$R(\phi_3,\phi_3^{\prime})$ & $\theta(\phi_3,\phi_3^{\prime})$ &   $\phi_3$ &  $\phi^{\prime}_3$ & $k_{couple}$\\
			\hline
			0.954 & -0.642$\pi$ &$\phi_{3AB}$  & $\phi_{3A}$  & 1.3 \\ 
			0,998 &  0.024$\pi$ &$\phi_{3AB}$  & $\phi_{3B}$  & 1.3\\ 
			0.925 & -0.292$\pi$ &$\phi_{3AB}$  & $\phi_{3A}$  & 0.6 \\ 
			0.920 &  0.373$\pi$ &$\phi_{3AB}$  & $\phi_{3B}$  & 0.6 		
		\end{tabular}
	\end{table}
}

These results confirm numerically that, when $K_{couple} = 1.3$, $\phi_{3AB(l=2.5)}$ generated by the light input $AB(l = 2.5)$ is better synchronized with  $\phi_{3B}$ generated by light input $B$ than with $\phi_{3A}$ generated by light input $A$. For $K_{couple} =0.6$ the differences are substantially smaller. 

One important question is if the results which we find for $l=2.5$ are an exception or if they are common in the parameter space. In the next subsection we present the results for this question using a more general analysis.

\section{Generalizing results}

\begin{figure}[H]
	\centering
	\includegraphics[scale=.7]{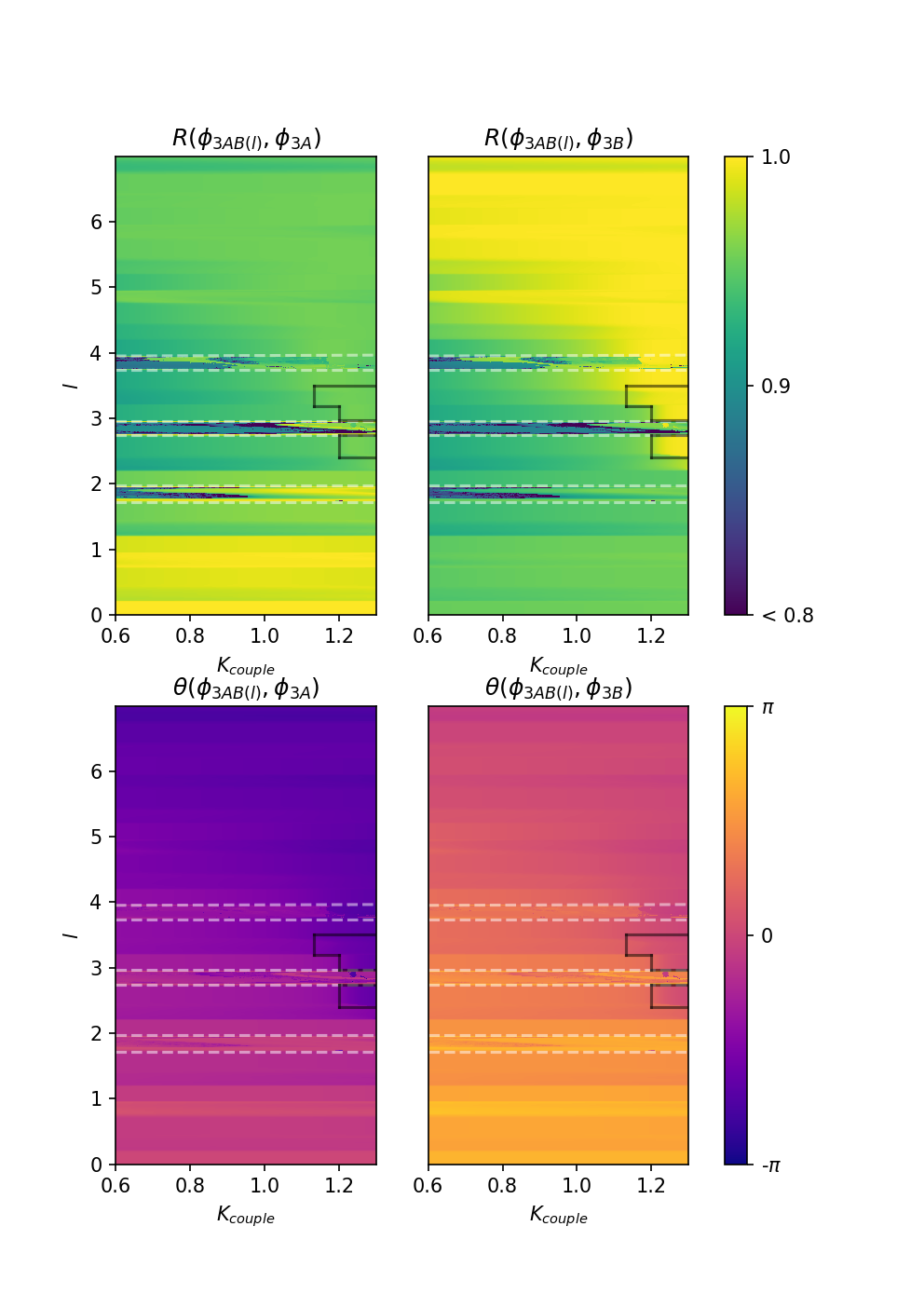}
	\caption{Parameter spaces related to synchronization between $\phi_{3AB(l)}$ and $\phi_{3A}$ or $\phi_{3B}$, which are, respectively, the phases of oscillator 3  when the light input is $AB(l)$, $A$, and $B$. $R(\phi_{3AB(l)},\phi_{3A})$[$R(\phi_{3AB(l)},\phi_{3B})$] is a synchronization measurement between  $\phi_{3AB(l)}$ and $\phi_{3A}$[$\phi_{3B}$]. $\theta(\phi_{3AB(l)},\phi_{3A})$ [$\theta(\phi_{3AB(l)},\phi_{3B})$]  is the mean angle between $\phi_{3AB(l)}$ and $\phi_{3A}$[$\phi_{3B}$]. $K_{couple}$ is the control parameter, $l$[7-$l$] is the number of days in the week that correspond to light input $B$[$A$] in the light input $AB(l)$.\label{fig:redt} }
\end{figure}

In order to have a more general view of the results, we  calculate the lengths $R(\phi_{3AB(l)},\phi_{3A})$,   $R(\phi_{3AB(l)},\phi_{3B})$ and the angles $\theta(\phi_{3AB(l)},\phi_{3A})$,  $\theta(\phi_{3AB(l)},\phi_{3B})$ of the first moments from the distributions  $z(\phi_{3AB(l)},\phi_{3A})$ and $z(\phi_{3AB(l)},\phi_{3B})$  for $l \in [0,7]$ with a step of 0.0125 day and $k_{couple} \in [0.6,1.3]$ with a step 0.00125. It is worth remembering here that the light input $AB(l)$ has $7-l$ days in the light input $A$ and $l$ days in the light input $B$, and that there is a shift of 8 h between $A$ and $B$. Therefore, we decide to vary $l$ to verify if the behavior that we find for $l=2.5$ and $K_{couple} = 1.3$ is usual or if is an exception. We are also interested to observe if other phenomena could appear in these other sets of parameters. In figure \ref{fig:redt} we present the results, in both cases in the axis x we have the coupling $k_{couple}$ and in axis y $l$.

In figure \ref{fig:redt} we basically see three types of behavior. The most common is the following:  if the time for ligtht stimus $A$ is higher than $B$, the output signal of the suprachiasmatic nucleus (SCN) $\phi_3$ is better synchronized to the output correspondent to light input $A$ than to the output correspondent to light input $B$ and vice-versa.
We can see this behavior comparing the parameters space on the figure \ref{fig:redt} on the left (related to light input $A$) to the parameters spaces on the right (related to light input $B$). For $l<3.5$, for the most values of $K_{couple}$, $R$ is lager on the left panel (related to light input $A$) than on the right panel (related to light input $B$), and $\theta$ is closer to 0 on the left panel than on the right panel, the regions where these properties do not occur will be explained later. Inversely, for $l > 3.5$, $R$ is larger on the right panel (related to light input $B$) than on the left panel (related to light input $A$), and $\theta$ is closer to 0 on the right panel than on the left panel.
In a smaller area of the parameter space, the behavior is the opposite of the common behavior.  This peculiar behavior is the same effect that we observe for the light input $AB(l = 2.5)$ with $k_{couple} = 1.3$. The regions where this behavior occurs are highlighted by continuous black borderlines.
\begin{figure}[H]
	\centering
	\includegraphics[scale=.7]{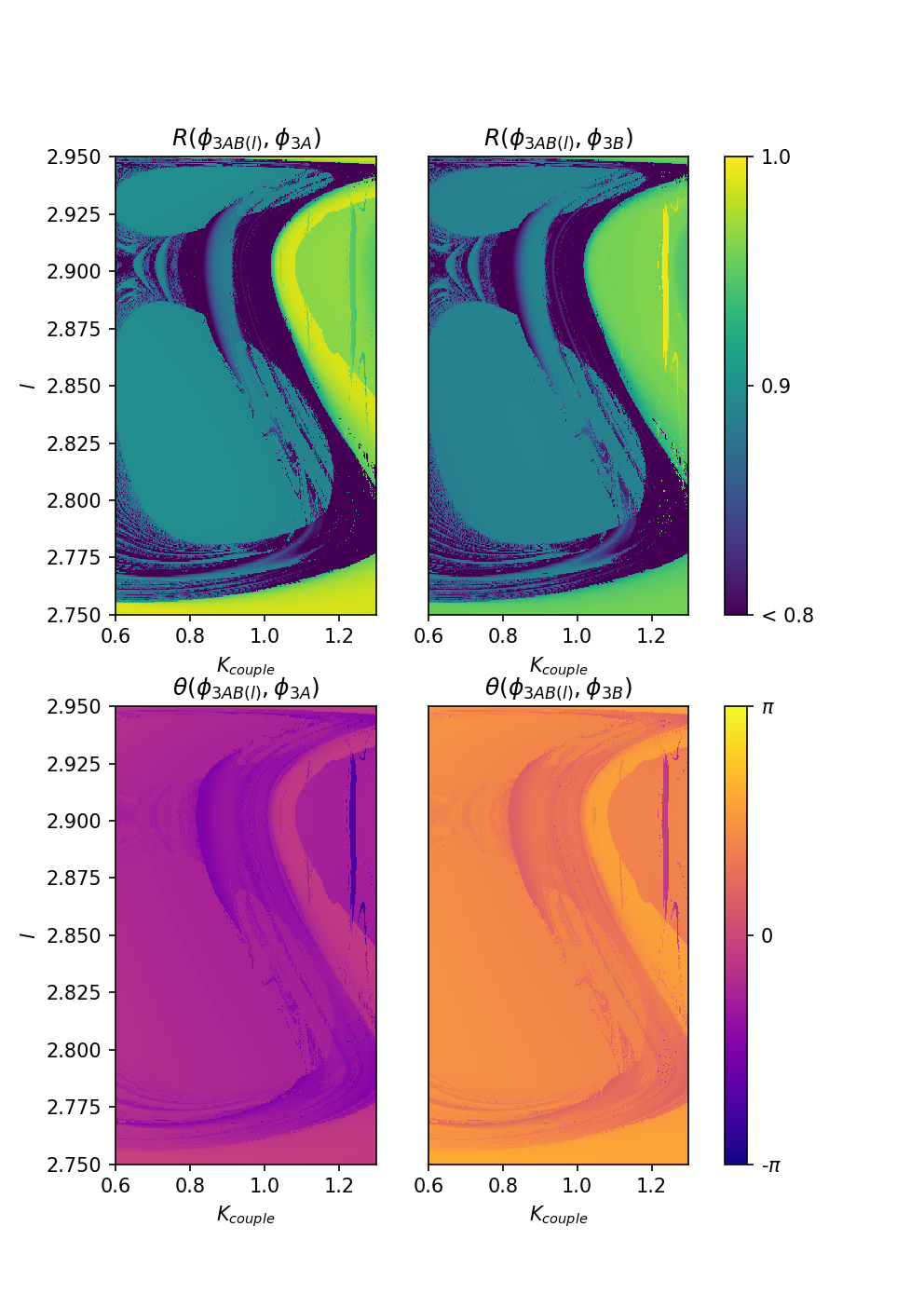}
	\caption{\label{fig:zoom1}Zoom of parameter spaces related to synchronization.}
\end{figure}
There are also three areas where neither the common behavior nor its opposite clearly occur. These regions are highlighted in dashed white borderlines  in figure \ref{fig:redt}, which we will call from now on regions $Y$. In these areas, the levels of synchronization are lower in general than in the rest of the parameter space and there is no much difference between the panels on the right and on the left. In each region $Y$ the colors patterns change abruptly in the colormaps of the lengths $R(\phi_{3AB(l)},\phi_{3A})$ and $R(\phi_{3AB(l)},\phi_{3B})$. In the colormaps of the angles $\theta(\phi_{3AB(l)},\phi_{3A})$ and $\theta(\phi_{3AB(l)},\phi_{3B})$ the variations are more subtle,   even so it is possible to notice that in regions $Y$ the color changes less regularly than in other regions. 

Aiming to better identify what is happening in regions $Y$, we simulate one of the regions $Y$ with more details. In this simulation $k_{couple}$ changes in a step of 0.001 and $l$ varies in a step of 0.005. We show the results in figure \ref{fig:zoom1}. We notice that region $Y$ actually presents an intricate structure, that is similar (not identical) in both left and right panels for $R$. The same structure appears for $\theta$, though the sign of the angles are opposite in the panels. While outside of the regions $Y$ the levels of coupling do not change much, or change smoothly with the changing of the $k_{couple}$ or $l$, in the regions $Y$ a slight change of parameter can change considerably the levels of the synchronization. Also only in regions $Y$ we find sets of the parameter space that assume the lowest values of synchronization in our scale.
\begin{figure}[H]
	\centering
	\includegraphics[scale=0.7]{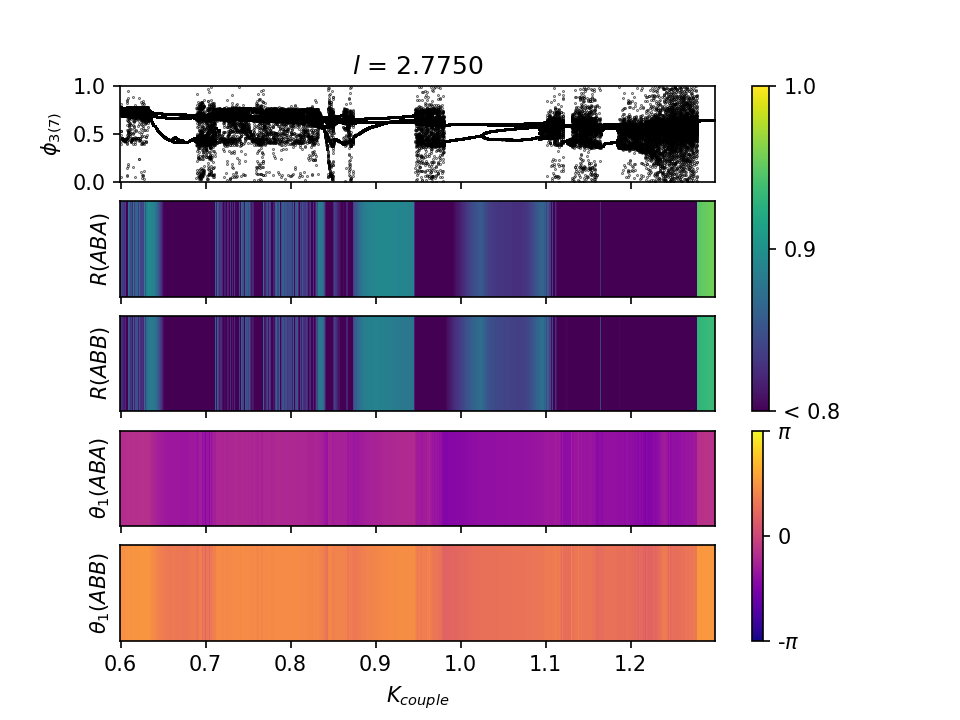}
	\caption{\label{fig:ex} $\phi_{3(7)}$ Value which $\phi_3$ assumes on the  Poincaré section in which we save values at each 7 days, abbreviation in figure  $R(\phi_{3AB(l)},\phi_{3A})$ as $R(ABA)$, $R(\phi_{3AB(l)},\phi_{3B})$ as $R(ABB)$, $\theta_1(\phi_{3AB(l)},\phi_{3A})$ as $\theta_1(ABA)$ and $\theta_1(\phi_{3AB(l)},\phi_{3B})$ as $\theta_1(ABB)$. }
	\label{fig:collage}
\end{figure}
The behavior in regions $Y$ is more complicated, as a typical example we show  the results for $l$ = 2.7750 in figure \ref{fig:ex}. For questions of space in this figure we shortened $R(\phi_{3AB(l)},\phi_{3A})$ as $R(ABA)$, $R(\phi_{3AB(l)},\phi_{3B})$ as $R(ABB)$, $\theta_1(\phi_{3AB(l)},\phi_{3A})$ as $\theta_1(ABA)$ and $\theta_1(\phi_{3AB(l)},\phi_{3B})$ as $\theta_1(ABB)$. We also show the value that $\phi_3$ assumes on the Poincaré section in which we save values at each 7 days $\phi_{3(7)}$. It is valid to remember here that the total period of the external forcing of this system is 7, since despite the changes during the week, at  each 7 days the cycles re-initiates. Comparing  $\phi_{3(7)}$ to $R(ABA)$, $R(ABB)$, $\theta_1(ABA)$ and $\theta_1(ABB)$  we  notice that the abrupt changes of colors are related to bifurcations. Also, we can see regions with period 2, 4 and chaotic windows, the last ones being the regions with lowest  values of $R(ABA)$ and $R(ABB)$. 

\begin{figure}[H]
	\centering
	\includegraphics[scale=0.7]{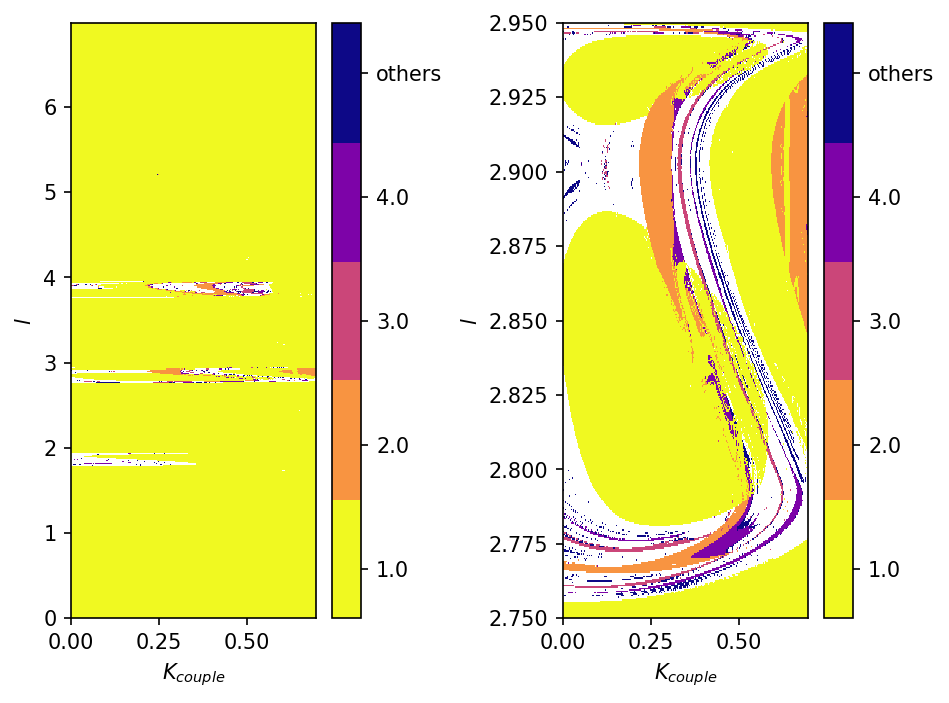}
	\caption{Parameter space of the period. The white regions are the regions that the algorithm could not find a period, either because it is too high or because the solution is chaotic. The other periods are indicated in the legend next to the panels. \label{fig:period}}
\end{figure}

\begin{figure}[H]
	\centering
	\includegraphics[scale=0.7]{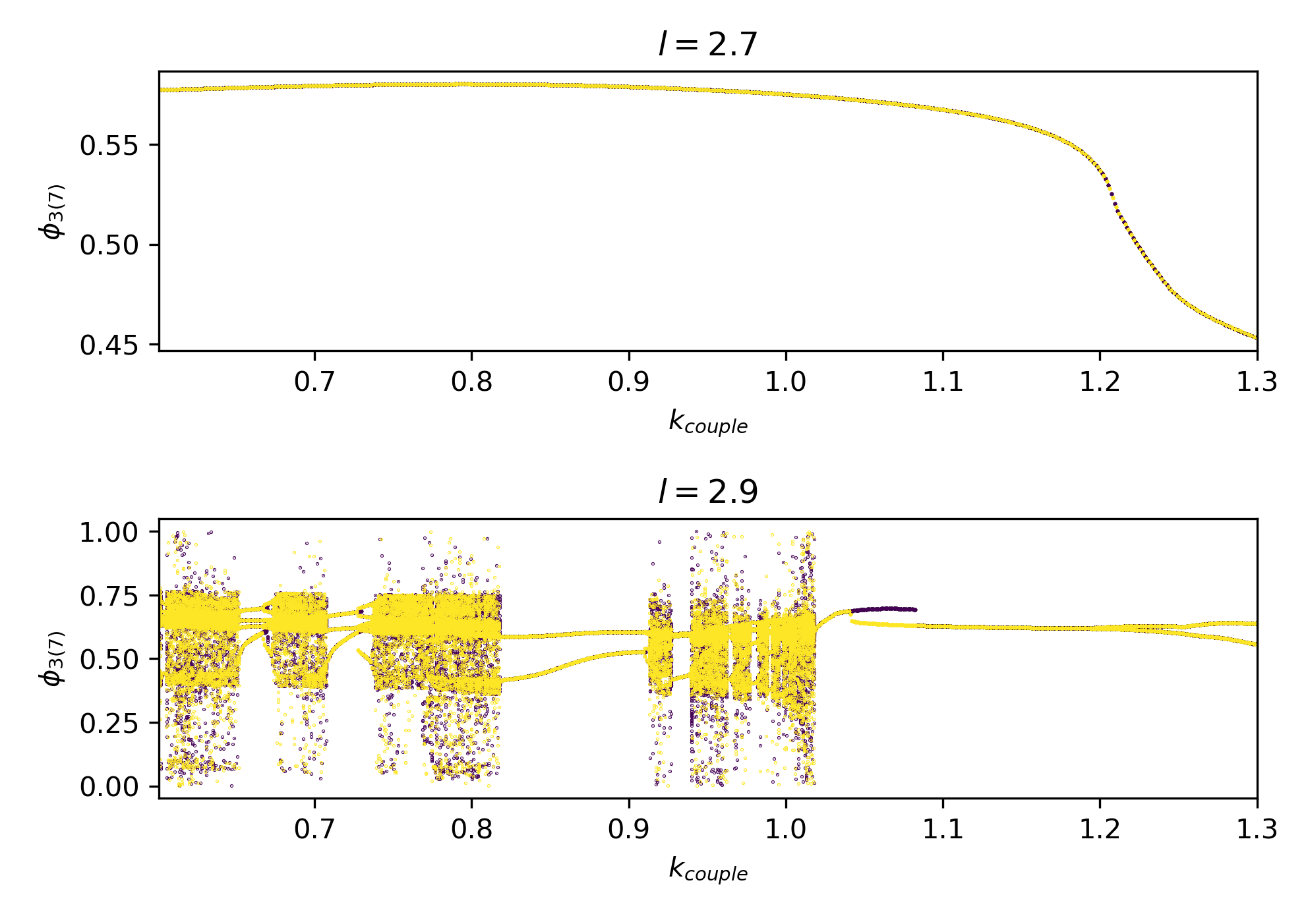}
	\caption{\label{fig:follow2} Value that $\phi_3$ assumes on the Poincaré section in which we take values at each seven days, for $l = 2.7$ and $l = 2.9$ following the attractor. In yellow simulations with $k_{couple}$ increasing and in purple decreasing. }
\end{figure}

\begin{figure}[H]
	\centering
	\includegraphics[scale=0.7]{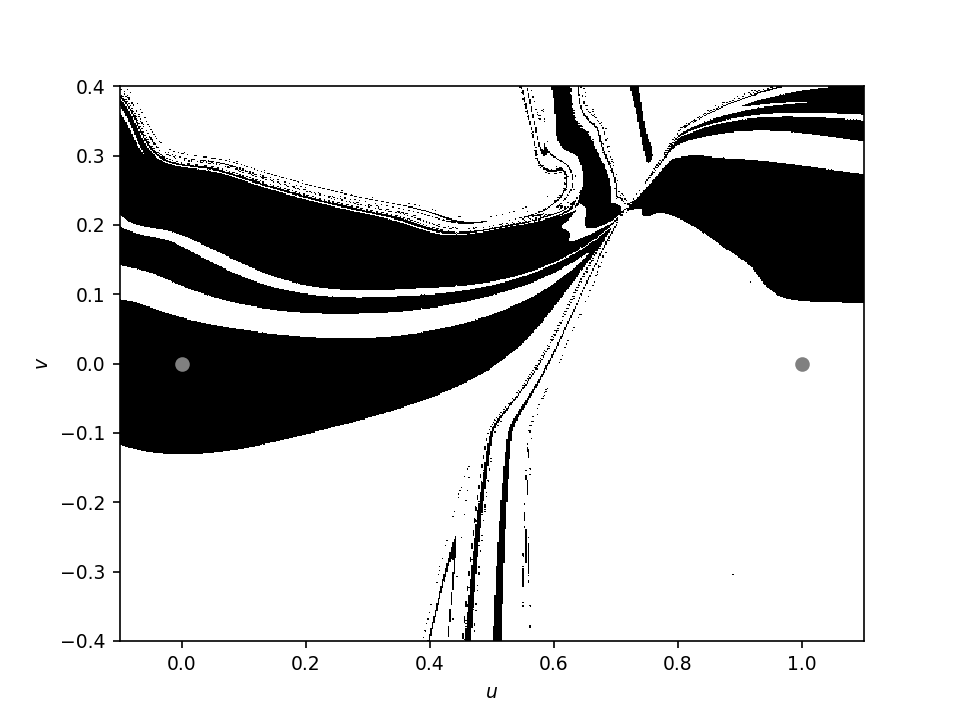}
	\caption{2D-projection of the basins of attraction for $K_{couple} = 1.06$ and $l = 2.9$. The black dots are the fixed points, the initial conditions in the black(white) region converge for the fixed point in the black(white) region.  We choose a projection in which both fixed points $s_0$ and $s_1$ belong to the plan, the axis $u$ is in the same direction of the vector $s_1 - s_0$, the axis $v$ is perpendicular to $u$ and is in the plan $\phi_1\_\phi_2$}
	\label{fig:bacia512}
\end{figure}

Observing the value which $\phi_3$ assumes on the Poincaré section (which we  construct taking one value at each seven days) in figure \ref{fig:ex} we already notice that in regions $Y$ there are parameters which correspond to solutions with period 1, with period 2, while other parameters present much higher period or even may be chaotic solutions. In order to analyze the periodicity of the solutions with more detail we present in figure \ref{fig:period} the period of the solutions firstly for the whole range $l \in [0,7]$ and $k_{couple} \in [0.6,1.3]$, secondly for the same range of the figure \ref{fig:zoom1} which we see that presents more intricate behavior. The color scale in figure indicates the values of the periods, the white regions are the regions that algorithm is not able to find a period (either because the period is too high, or because the solution is chaotic). In the majority of the space we  find period 1, but in the three regions that we find more complicated behavior other values of periods and white regions appear. In the second figure we also notice that the shape of the colorplot of the periods and of the quantities  $R(\phi_{3AB(l)},\phi_{3A})$,  $R(\phi_{3AB(l)},\phi_{3B})$ have similar shapes, which is reasonable, since when we look at one line in figure \ref{fig:ex} the bifurcations and the abrupt changes of colors coincide. 
Another point to analyze is if there is bistability. To analyze that we simulate the system following the attractor, that is, using the final result of the simulation of one parameter as initial condition for the simulation for the next parameter. In these simulations the value of $l$ is fixed and the value of $k_{couple}$ is changed. We find out that that for some values of $l$ the attractor is the same when simulation is done with 
$k_{couple}$ increasing or decreasing. In figure \ref{fig:follow2} we show two
examples, for $l = 2.7$ nothing changes if the simulation is done with $K_{couple}$ increasing
or decreasing, for $l = 2.9$ though there is a region in which we see bistability.
As a way to visualize better this bistability we fixed the parameters $k_{couple} =1.06$, $l = 2.9$ since from the figure \ref{fig:follow2} we know that these parameters belong to a region where bistability occurs. In the figure   \ref{fig:bacia512} we present a 2D projection of the basins of attraction. The gray dots are the fixed points, the initial conditions in the black (white) region converge for the fixed point in the black (white) region.  We choose a projection in which both fixed points $s_0$ and $s_1$ belong to the plan, the axis $u$ is in the same direction of the vector $s_1 - s_0$, the axis $v$ is perpendicular to $u$ and is in the plan $\phi_1\_\phi_2$.

We also analyze the cases for different values of shifts between light inputs $A$ and $B$, including not integer values of shifts. For lower values of shift we only find the common behavior, but as the shift increases we find the opposite behavior as well as areas with higher period and chaotic solutions. We present these results in the supplementary material. From those simulations, we conclude that different behaviors from the common one occur when the shift between the light inputs $A$ and $B$ is high enough.

	We also simulate the system for $k_{couple} > 1.3$ even though these exceeds the limits used in \cite{yamaguchi_mice_2013}.  More regions with solutions of higher period and chaos appear. Also we see an increase in the area of the region in which the opposite behavior of the common occurs, but the area in which this effect  stands out more is visible in figures presented here. We add the figures for $K_{couple} > 1.3$ in the supplementary material.

\section{Conclusions}

We emulate the effect of the social jetlag in the model introduced in \cite{yamaguchi_mice_2013}, changing the light input by a light input sequence which represents a person's routine who suffers from social jetlag. The light input that we considered has a 7-days period and it is composed by a succession of two light inputs $A$ and $B$. Both light inputs $A$ and $B$ are composed of 12h of light and 12h of dark, the only difference between them being a shift. The shift in the figures presented here is of 8h, but we also analyzed other cases that we show in the supplementary material.

In reference \cite{yamaguchi_mice_2013} the authors found a simple relation, namely, the lower the coupling parameter the faster the recovery of the jetlag (shift in the light input). In our case, every week the system is subjected to a shift on the light input, so we made simulations to measure if the system is better synchronized to light input $A$ (light input before the weekly shift) or to light input $B$ (light input after the weekly shift). In order to improve our analyses we also determined the period of the solutions on a Poincaré Section that we took one value per week. There was not a simple analogous of the result presented in \cite{yamaguchi_mice_2013}. However, we were able to find that there are regions in the parameter space where the output of the system is better synchronized in relation to light input $A$ or $B$. We also highlighted the regions of the parameter spaces that differently from the rest have higher period and even chaotic solutions. 

The most common behavior that we found was: if the individual spends more time under a light stimulus $A$($B$) than $B$($A$) during the week, the output of the system is better synchronized to light input $A$($B$). However, in a small parameter space area, we also encountered the opposite behavior. It occurs due to the high nonlinearity of the model. There is a minimal shift between the light inputs $A$ and $B$ which is needed for this behavior to appear. 

We could not clearly identify either the common behavior or its opposite in some regions in the parameter space. In these areas small changes of parameters could lead to large variations on the levels of synchronizations. In these regions  the lowest values of synchronization level of our simulation were reached. Almost  the whole parameter space had solutions of period 1, the exceptions were exactly these regions where we saw solutions of higher periods or even chaotic solutions. Furthermore, for some parameters there is bistability. Again this type of behavior does not occured if the shift between the two light inputs is too short. 

 It would be very informative to find if individuals under light inputs  with systematic shifts, similar to those considered, present high period or chaotic output signals.. Although it is possible to collect some data about people who suffer from social jetlag and analyze it, an experiment using mice would be very interesting, because then one could control with precision the light input as well as interfere in the coupling of the neurons. Also it could be constructive to analyze how other models behave when subjected to the type of light input that we studied in this work.  

\section{Acknowledgements}

The authors acknowledge the financial support from the
Brazilian Federal Agency CNPq,  Grant No. 302665/2017-0,  and the São Paulo Research Foundation (FAPESP, Brazil), under
Grant Nos. 2018/03211–6 and 2018/22140-2.

\bibliography{mybibfilecorrected}

\end{document}


\begin{frontmatter}

\title{Supplementary Material}

\author{Flávia M. Ruziska\fnref{myfootnote}\corref{mycorrespondingauthor}\fnref{myfootnote}}
\cortext[mycorrespondingauthor]{Corresponding author}
\ead{flaviamayumi.rh@gmail.com}
\author{Iberê L. Caldas \fnref{myfootnote}}
\address{Instituto de Física, Universidade de S\~{a}o Paulo, Rua do Mat\~{a}o, 1371, 05508-090 S\~{a}o Paulo, SP, Brazil}



%
%
%

\end{frontmatter}

\section{Simulations including $K_{couple} > 1.3$}

\begin{figure}[H]
	\centering
	\includegraphics[scale=0.6]{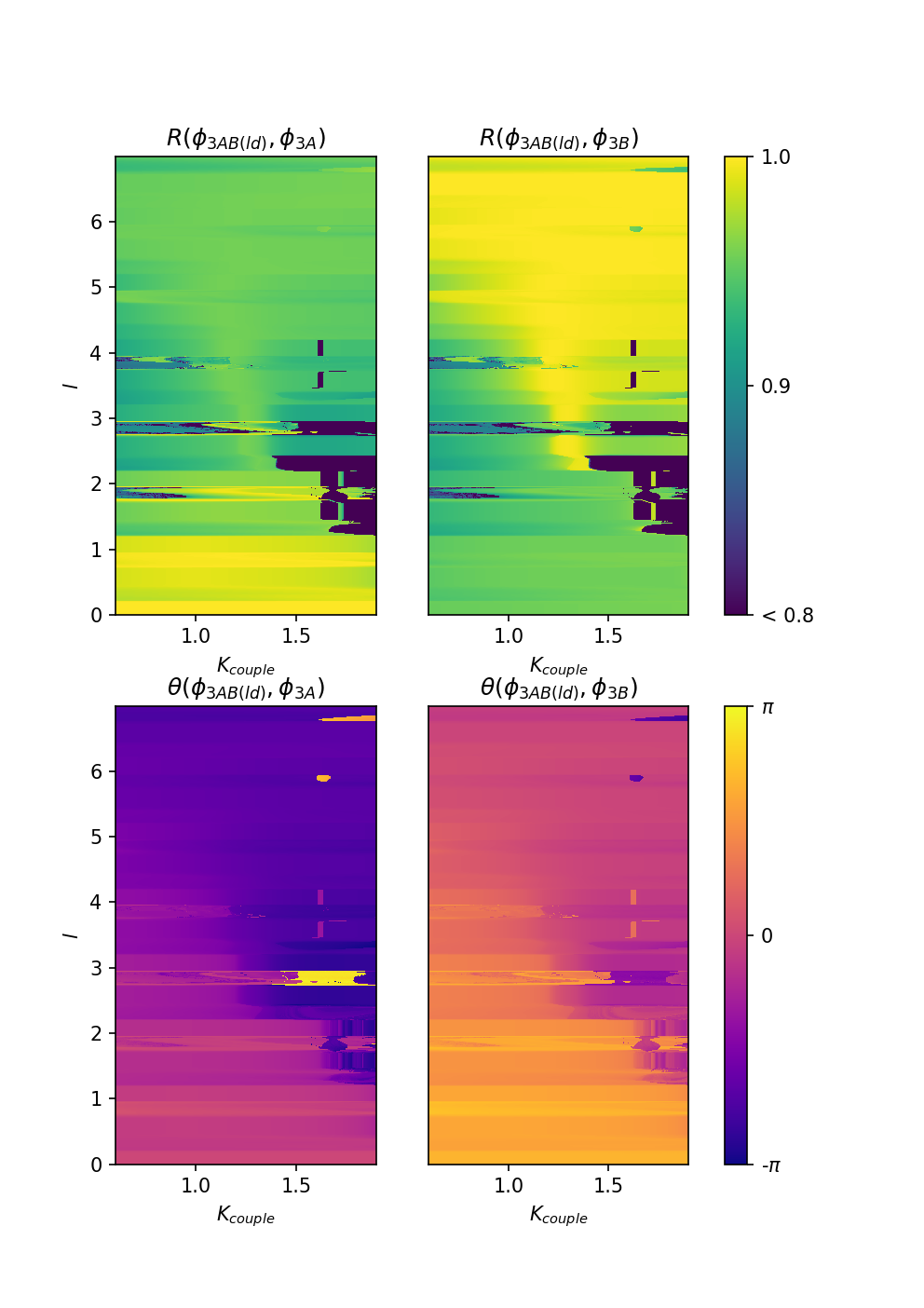}
	\caption{Color representations of $R(\phi_{3AB(l)},\phi_{3A})$, $R(\phi_{3AB(l)},\phi_{3B})$,  $\theta_1(\phi_{3AB(l)},\phi_{3A})$ and $\theta_1(\phi_{3AB(l)},\phi_{3B})$ as a function of $l$ and $K_{couple}$ for shift $= 8h$, and $K_{couple} \in [0.6,1.9].$}
	\label{fig:resultsexpanded}
\end{figure}

In all simulations presented in the article we varied the parameter $K_{couple}$ in the range [0.6,1.3]
 that are the extreme values considered in \cite{yamaguchi_mice_2013}. Nevertheless, since that the less common behavior (better synchronization to the light input that occurs less time on the week) occurs for a set the parameters in which the $K_{couple}$ is close to 1.3, we considered important simulate the system for higher values of $K_{couple}$.  Comparing the panels on the left and on the right in figure \ref{fig:resultsexpanded} we notice that the region where the less common behavior occurs increases. However the regions where the effect is more pronounced is around $K_{couple} = 1.3$, increasing the coupling do not always increase the effect.

\begin{figure}[H]
	\centering
	\includegraphics[scale=0.6]{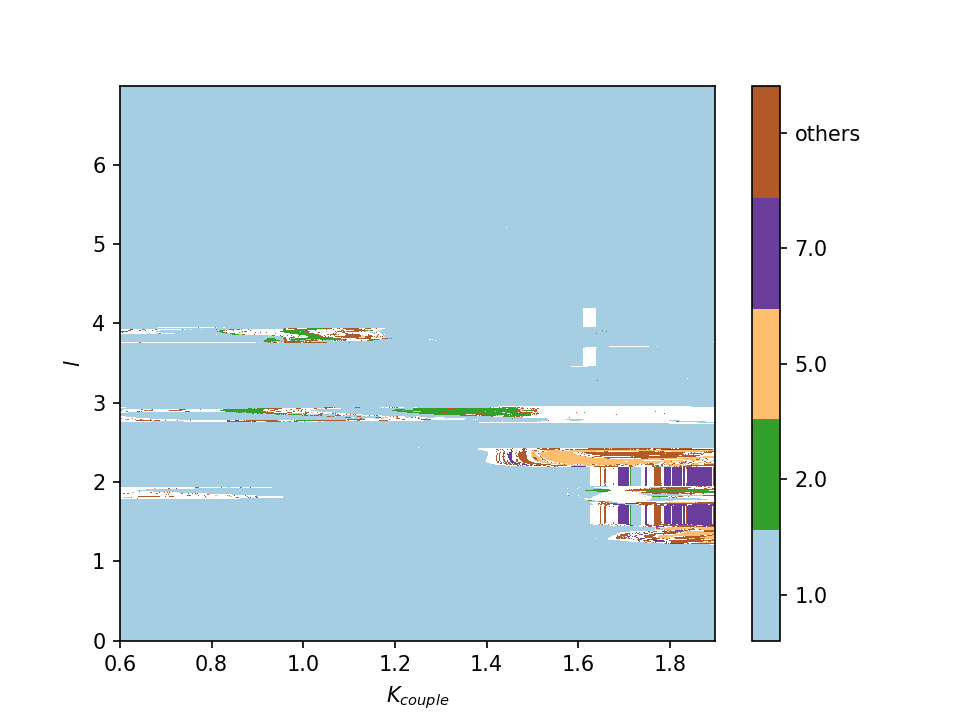}
	\caption{Parameter space of the period for shift $= 8h$, and $K_{couple} \in [0.6,1.9].$. The white areas represent aperiodic solutions. The other periods are indicated in the scale next to the panels.}
	\label{fig:preiodresultsexpanded}
\end{figure}

 On the other hand we find out that for larger values of $K_{couple}$ more regions with solutions with high periods or even chaotic appear. We can see these regions in figure \ref{fig:preiodresultsexpanded}. In the synchronization parameter spaces the same areas are the regions of lower values of $R$ in both panels on the right and on the left, in that case the levels of synchronization are lower in relation to both light inputs $A$ and $B$. 
  
\section{Simulations for different values of shifts between the light input A and B}

In the article we presented the results for a light input composed by sequence of two light inputs $A$ and $B$ with a shift of 8h between them. Here we present the results when we consider other values of shift. As we mentioned in the article we concluded that for lower values of shift, such as shift $= 6h$ the only type of synchronization was to the light input that occurs longer on the week and all solutions have period 1 (on a Poincaré section for one value per week). 

As the shift increases the other behaviors described in the article appear. There is the opposite of the common behavior (better synchronization to the light input that occurs less time on the week), and also there are regions where neither the most common behavior nor its opposite clearly occur. In these last regions we observe in the parameter space of the periods that the solutions have higher periods or even are chaotic. 

In the following we present the synchronization parameter spaces as well the parameter spaces of the period for some values of shift. We do not present the parameter space for shift $= 6h$ because in that case we find period 1 for all parameters. 

\begin{figure}[H]
	\centering
	\includegraphics[scale=0.6]{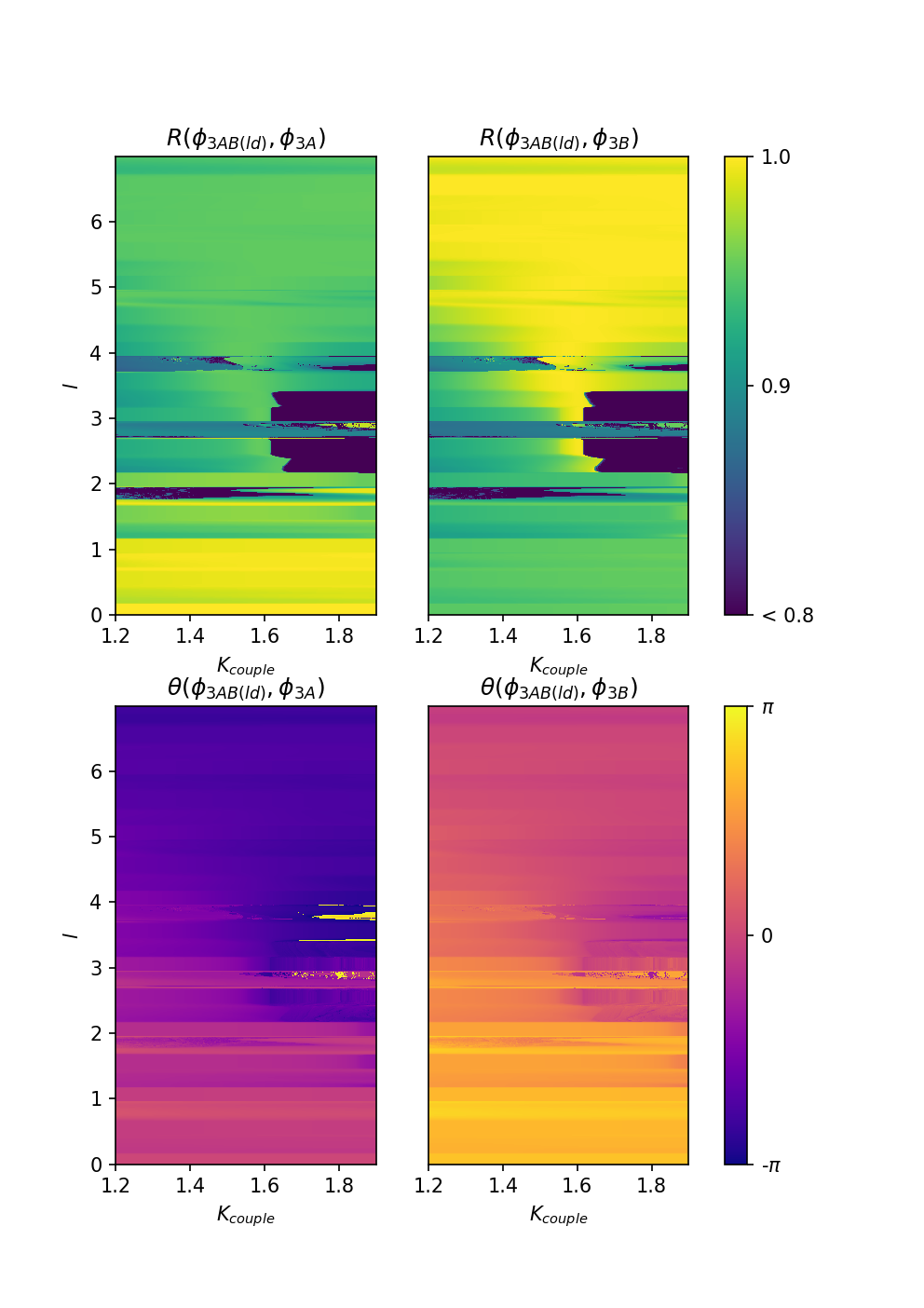}
	\caption{Color representations of $R(\phi_{3AB(l)},\phi_{3A})$, $R(\phi_{3AB(l)},\phi_{3B})$,  $\theta_1(\phi_{3AB(l)},\phi_{3A})$ and $\theta_1(\phi_{3AB(l)},\phi_{3B})$ as a function of $l$ and $K_{couple}$ for shift  = $9h$ }
	\label{fig:lag9h}
\end{figure}

\begin{figure}{H}
	\centering
	\includegraphics[scale=0.6]{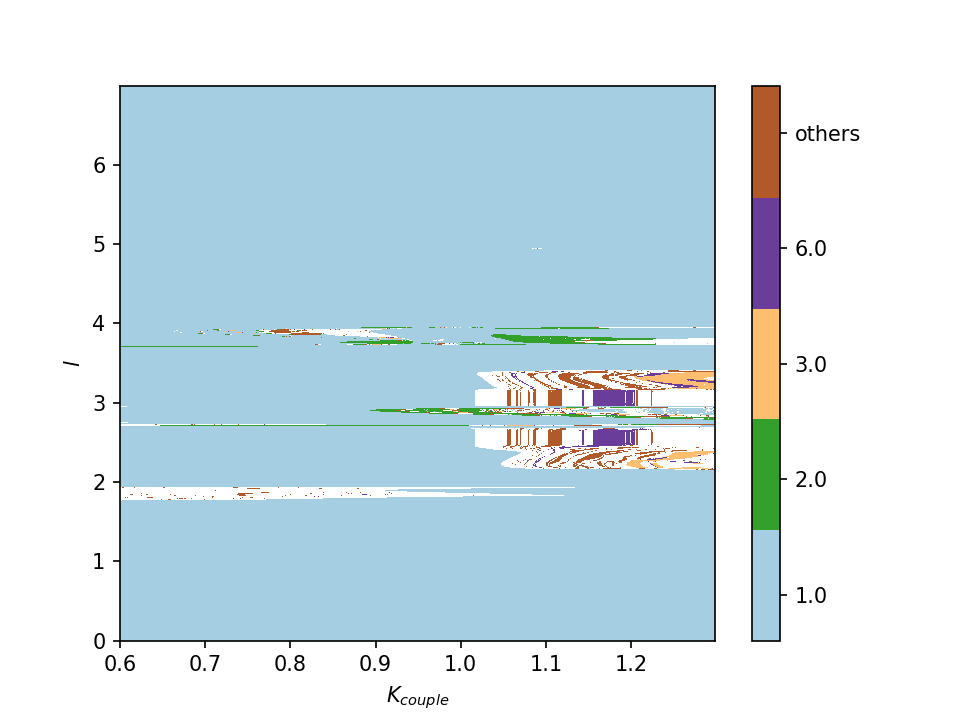}
	\caption{Parameter space of the period for shift $= 9h$, and $K_{couple} \in [0.6,1.3].$. The white areas represent aperiodic solutions. The other periods are indicated in the scale next to the panels.}
	\label{fig:period_lag9h}
\end{figure}

%
\clearpage

\begin{figure}[H]
	\centering
	\includegraphics[scale=0.6]{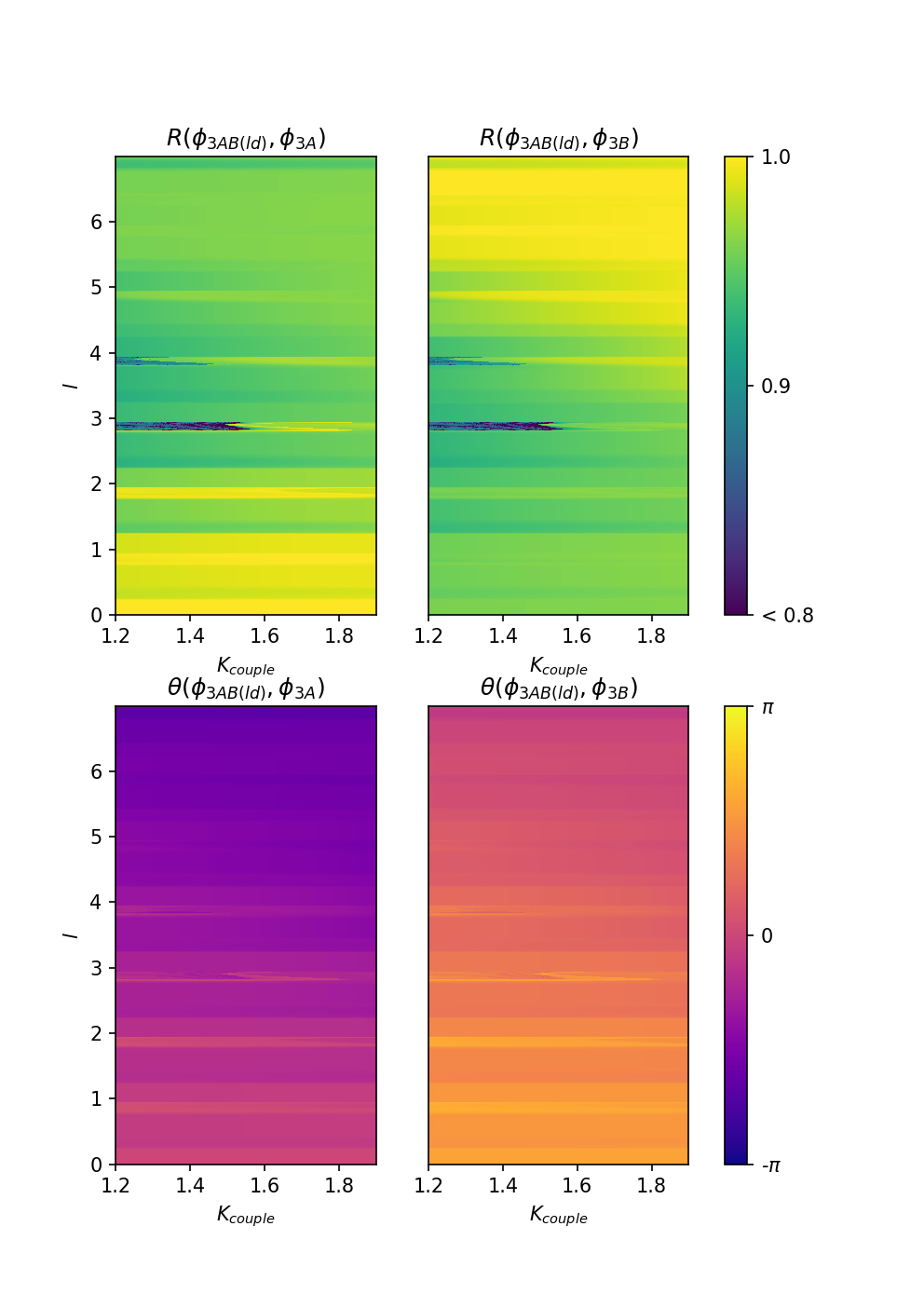}
	\caption{Color representations of $R(\phi_{3AB(l)},\phi_{3A})$, $R(\phi_{3AB(l)},\phi_{3B})$,  $\theta_1(\phi_{3AB(l)},\phi_{3A})$ and $\theta_1(\phi_{3AB(l)},\phi_{3B})$ as a function of $l$ and $K_{couple}$ for shift $= 7h$}
	\label{fig:lag7h}
\end{figure}

\begin{figure}[H]
	\centering
	\includegraphics[scale=0.6]{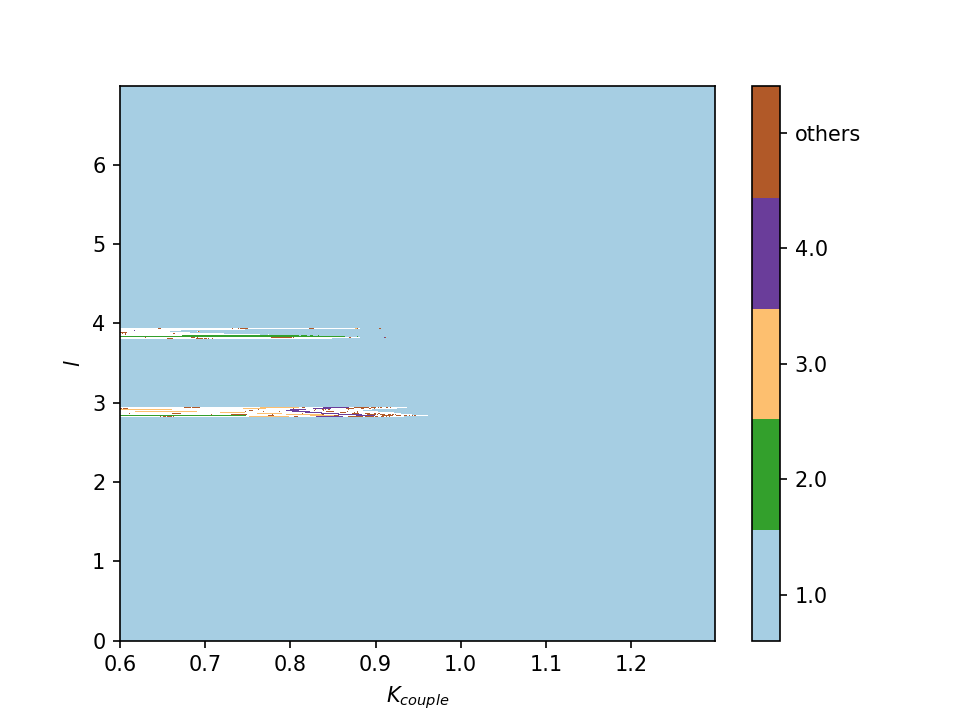}
	\caption{Parameter space of the period for shift $= 7h$, and $K_{couple} \in [0.6,1.3].$. The white areas represent aperiodic solutions. The other periods are indicated in the scale next to the panels.}
	\label{fig:period_lag7h}
\end{figure}

\begin{figure}[H]
	\centering
	\includegraphics[scale=0.6]{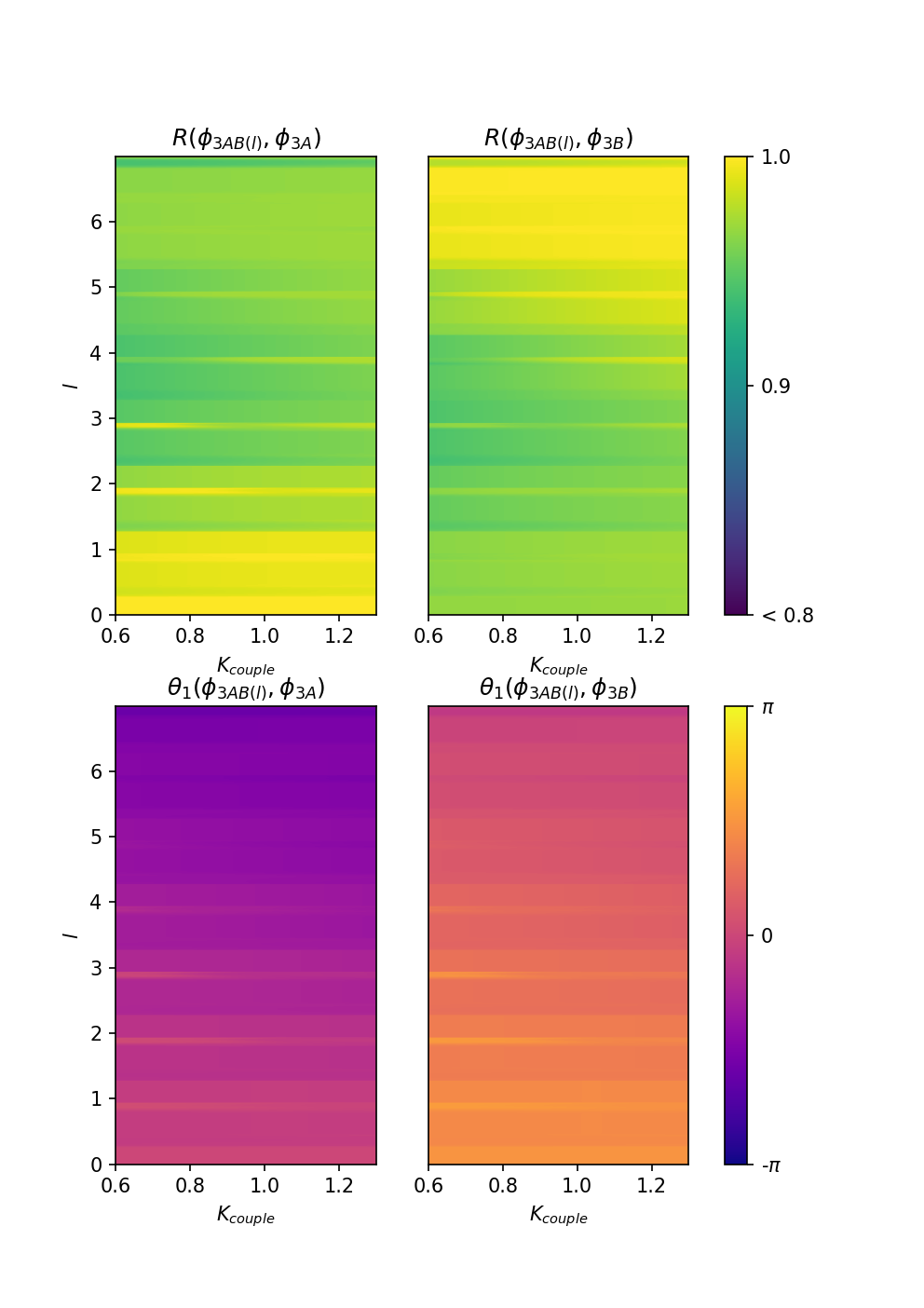}
	\caption{Color representations of $R(\phi_{3AB(l)},\phi_{3A})$, $R(\phi_{3AB(l)},\phi_{3B})$,  $\theta_1(\phi_{3AB(l)},\phi_{3A})$ and $\theta_1(\phi_{3AB(l)},\phi_{3B})$ as a function of $l$ and $K_{couple}$ for shift $= 6h$}
	\label{fig:lag6h}
\end{figure}

\bibliography{mybibfilecorrected}